\documentclass[10pt,leqno]{amsart}

\usepackage{graphicx}
\usepackage{indentfirst,csquotes}
\usepackage{hyperref}

\usepackage{xcolor,paralist,titlesec,fancyhdr,etoolbox}
\usepackage[symbol]{footmisc}  

\titleformat{\section}[display]{\normalfont\huge\bfseries\centering}{\centering\chaptertitlename\thechapter}{10pt}{\Large}
\titlespacing*{\section}{0pt}{0ex}{0ex}

\newcommand*\taccari{1}
\newcommand*\ovadia{2}
\newcommand*\wang{3}
\newcommand*\kahana{4}
\newcommand*\chen{1}
\newcommand*\jimack{3}

\title{Understanding the Efficacy of U-Net \& Vision Transformer for Groundwater Numerical Modelling}

\author{Maria Luisa Taccari\textsuperscript{\taccari}, Oded Ovadia\textsuperscript{\ovadia}, He Wang\textsuperscript{\wang}, Adar Kahana\textsuperscript{\kahana}, Xiaohui Chen\textsuperscript{\chen}, Peter K. Jimack\textsuperscript{\jimack}}

\date{\today}

\begin{document}

\maketitle

\footnotetext[\taccari]{\textsuperscript{\taccari}School of Civil Engineering, University of Leeds, Leeds, UK, Email: cnmlt@leeds.ac.uk.}
\footnotetext[\ovadia]{\textsuperscript{\ovadia}Department of Applied Mathematics, Tel-Aviv University, Tel-Aviv, Israel.}
\footnotetext[\wang]{\textsuperscript{\wang}School of Computing, University of Leeds, Leeds, UK.}
\footnotetext[\kahana]{\textsuperscript{\kahana}Department of Applied Mathematics, Tel-Aviv University, Tel-Aviv, Israel.}
\footnotetext[\chen]{\textsuperscript{\chen}School of Civil Engineering, University of Leeds, Leeds, UK.}
\footnotetext[\jimack]{\textsuperscript{\jimack}School of Computing, University of Leeds, Leeds, UK.}

\maketitle

\begin{abstract}
    This paper presents a comprehensive comparison of various machine learning models, namely U-Net \cite{ronneberger2015unet}, U-Net integrated with Vision Transformers (ViT) \cite{dosovitskiy2021image}, and Fourier Neural Operator (FNO) \cite{li2020fourier}, for time-dependent forward modelling in groundwater systems. Through testing on synthetic datasets, it is demonstrated that U-Net and U-Net + ViT models outperform FNO in accuracy and efficiency, especially in sparse data scenarios. These findings underscore the potential of U-Net-based models for groundwater modelling in real-world applications where data scarcity is prevalent.
\end{abstract}

\section*{Introduction}
    Groundwater numerical models, such as MODFLOW \cite{modflow}, are crucial for water resource management, although they are computationally demanding. To alleviate this, surrogate modelling through data-driven methods offers efficient approximations of these complex numerical techniques.

    Neural Operators \cite{deeponet, seidman2022nomad}, particularly the Fourier Neural Operator (FNO) \cite{li2020fourier}, have been at the forefront of recent advances, having shown potential to approximate arbitrary continuous functions. 
    However, the computational demand of FNO is particularly high during training phase while these neural operators require architectural enhancements to deliver promising results in subsurface problems \cite{wen2022ufno, jiang2023fouriermionet}. This is evident in the work of Wen et al. \cite{wen2022ufno}, where the integration of FNO with U-Net architecture showed improved accuracy, speed, and data efficiency in multiphase flow problems. However, Gupta and Brandstetter's work \cite{gupta2022multispatiotemporalscale}, showing that U-Net outperforms FNOs across various fluid mechanics problems, raises a question about the necessity of neural operators when the vanilla U-Net architecture already exhibits remarkable performance. 
    
    Recently, transformers \cite{vaswani2017attention} have seen considerable success in various fields, including physical systems \cite{cao2021choose, li2023transformer}, for which the datasets are typically smaller compared to other domains. Only one study explores the use of transformers  in groundwater modeling  \cite{francestudy}, demonstrating that the models were outperformed by both GRU and LSTM models to predict groundwater levels  across various stations in France with meteorological and hydrological data. 
    
    Finally, the integration of U-Net with Transformers, as exemplified in studies like TransUNet \cite{transunet} and ViTO \cite{vito}, has demonstrated their utility across a broad range of applications, particularly in the field of medical image segmentation and operator learning for inverse PDE problems. Yet, the applicability of these combinations in addressing time-dependent forward problems, real-world data scenarios, and in situations with sparse data, remain areas yet to be fully explored.

    Several studies, such as the one by Brakenhoff et al. \cite{brakenhoff, francestudy}, primarily focus on individual time series when analysing the impact of various hydrological stressors, including pumping rates, precipitation excess, and river stage variations, on groundwater levels of individual monitoring wells. While this approach provides valuable insights, it does not account for spatial correlations, thereby limiting its use to existing time series or monitoring wells. Similarly, previous comparisons have been predominantly limited to specific models like LSTM, CNNs and NARX in the context of groundwater level forecasting \cite{Wunsch_comparison}, leaving room for broader explorations.
    
    In this paper, we present a comprehensive comparison among models—specifically U-Net, U-Net integrated with Vision Transformers (U-Net+ViT), and Fourier Neural Operator (FNO)—for their efficacy in modeling time-dependent forward and inverse problems in groundwater systems. We test our model extensively on synthetic datasets, simulating conditions from the Overbetuwe region in the Netherlands, including sparse data scenarios. We show that both U-Net and U-Net+ViT are particularly well-suited to these important sparse data scenarios, with the addition of the Transformer providing enhanced predictive capability in many cases.

\section*{Methodology}
    \subsection*{Example of study site and data} \label{2.1}
\begin{figure}[ht]
    \vskip 0.2in
    \begin{center}
    \centerline{\includegraphics[width=0.5\columnwidth]{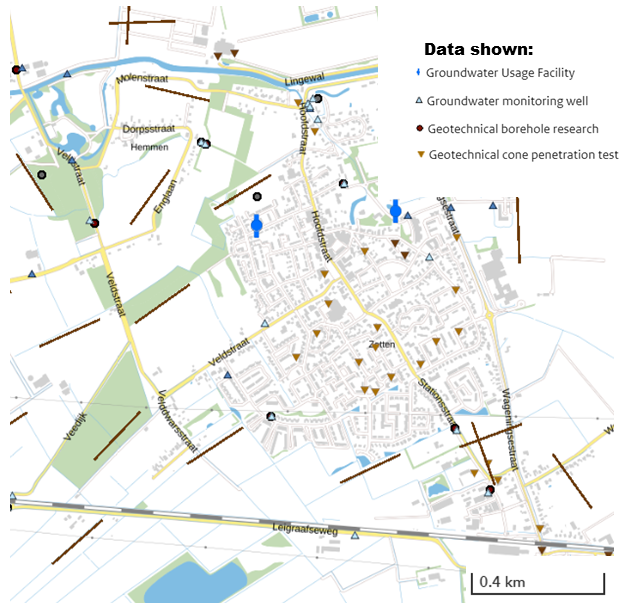}}
    \caption{A representation of the target study area located in the polder region of Overbetuwe, Netherlands. \cite{DINO_loket}.}
    \label{fig:loc-example}
    \end{center}
    \vskip -0.2in
    \end{figure}

    This subsection provides context and rationale for our study via an example case study based upon the polder region of Overbetuwe in the Netherlands (Figure~\ref{fig:loc-example}). This region showcases the characteristic Dutch system of water management where the area is divided into several polders in a mix of agriculture, nature, and urban environments. Alongside its sparse data and heterogeneous soil, these unique characteristics underscore the inherent complexities of water management in similar settings, making this dataset a suitable choice for our research. The subsoil is primarily composed of clay and sandy clay, with soil properties being determined via borehole and cone penetration tests. The study area features numerous observation wells for monitoring groundwater heads while well fields (indicated as groundwater usage facilities in the figure) are utilized for the extraction of drinking water. The work of Brakenhoff et al. \cite{brakenhoff} considers a dataset consisting of 250 head time series, with daily recordings starting from the year 1990 and drawdown attributed to the extraction from up to four well fields. 
    
\begin{figure}[ht]
    \vskip 0.2in
    \begin{center}
    \centerline{\includegraphics[width=1\columnwidth]{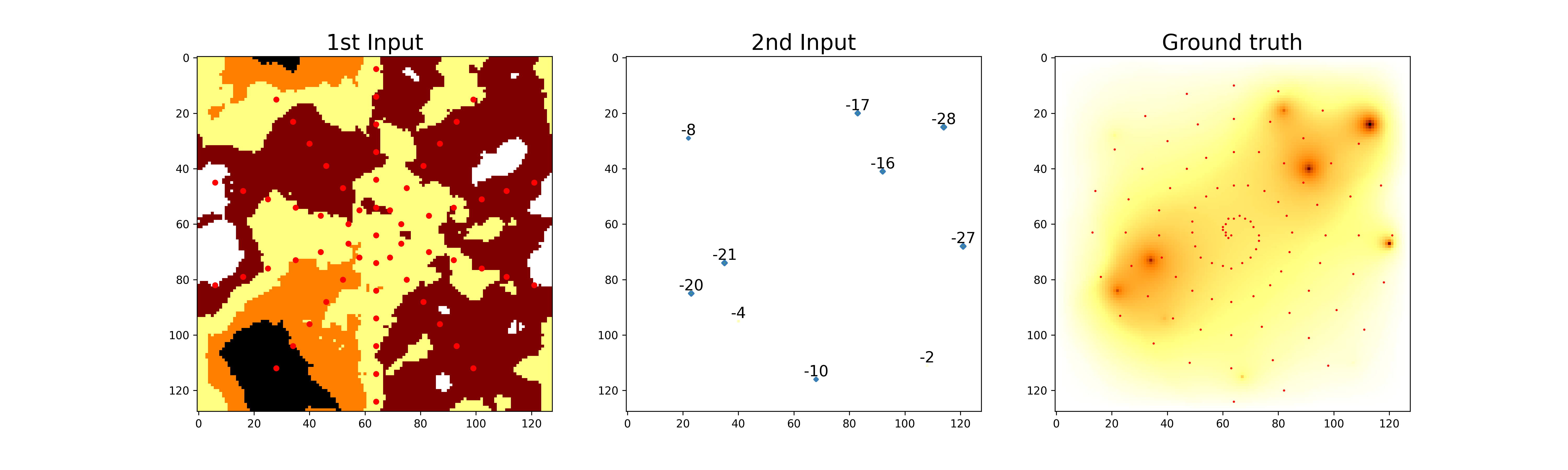}}
    \caption{An test sample of the high-fidelity dataset constructed using MODFLOW. The first column showcases the heterogeneous hydraulic conductivity, the second column presents the randomly positioned pumping wells, and the third column depicts the resulting hydraulic head. The red dots (column 1 and 3) represent a selection of data points—following radial and helical patterns, respectively—emulating the sparse observations in real-world scenarios.}
    \label{fig:input-example}
    \end{center}
    \end{figure}

    For the purposes of this study, we employ synthetic data to validate the proposed methodology, with the intention to subsequently apply the validated method to the real-world data of the Overbetuwe region. Figure~\ref{fig:input-example} represents a sample of the high-fidelity labeled dataset, which is constructed using the U.S. Geological Survey (USGS) finite-difference flow model, MODFLOW. The model is composed of a single-layer representation of a confined aquifer with a $128\times128$ grid. 
    The aquifer's heterogeneity is reflected through varying horizontal hydraulic conductivity within the bounds $k \in[0.1, 0.5]$ m/d. \textcolor{black}{The hydraulic conductivity fields in our study are created using random fields which are then thresholded to delineate different classes.}
    A maximum of ten pumping wells are extracting water with variable rates in the range $Q \in[0, 30]$ m$^3$/d over a simulation period of $T = 10$ days. The pumping wells are located in random locations which vary for each sample. The boundary conditions are delineated as Dirichlet, with the head equal to zero, mimicking a polder encircled by ditches where a stable water level is maintained through a comprehensive network of pumping stations.
    
    The datasets consist of $N_{train} = 5000$ training instances and $N_{test} = 1000$ testing instances. To mirror the inherent sparsity of real-world data, a data selection strategy is adopted for the test dataset. The locations of the boreholes for estimating the hydraulic conductivity are chosen following a radial distribution pattern, and a helical pattern is used for the wells monitoring hydraulic head (Figure ~\ref{fig:input-example}).
        
    \subsection*{Architectures}
    
    The architectures of the three models under comparison in this study encompass the U-Net structure, a U-Net with  attention mechanism in the bottleneck, and the Fourier neural operator (FNO).

    The U-Net architecture is designed with an encoder-decoder structure where the decoder receives the upsampled feature map, which is then concatenated with the corresponding feature map from the encoder through a skip connection. \textcolor{black}{Detailed diagrams of the the U-Net encoder and decoder can be found in Figures \ref{fig:encoder} and \ref{fig:decoder} in Appendix A.} The encoder consists of three bottleneck blocks, where each block utilizes three layers of Conv2d, Instance Normalization, and GELU activation to extract spatial features. These blocks increase the number of channels by a factor of 2 and perform downsampling with a stride of 2. The decoder is composed of a series of upsampling blocks, where each block consists of a bilinear upsampling operation (Upsample), followed by a double convolution operation. Each convolution within the decoderis followed by Instance Normalization and GELU activation function. The bottleneck consists on a single convolutional layer. In the time-dependent scenario, the time series data of the historical pumping rates is processed through two layers of feed-forward neural network (FNN) prior to being concatenated to the input for the latent space representation (Figure \ref{fig:encoder}).

    The second model, here called UNet+ViT, employs the Vision Transformer (ViT) \cite{dosovitskiy2021image}, in the latent space representation of the U-Net, as per implementation of TransUNet \cite{transunet} and ViTO \cite{vito}. The input is tokenized into a sequence of flattened 2D patches, each of size 1×1. Positional information is retained by employing trainable convolutional projection to learn and add specific position embeddings to the patch embeddings. The structure of the Transformer includes $L$ blocks, with each block comprising Multi-Head Attention (MSA) and FNN. This configuration involves the use of 2 blocks, each with 2 Multihead Self-Attentions, and a FNN composed of 128 neurons. \textcolor{black}{For a more detailed visualization of the Vision Transformer, attention block, and multihead attention, please refer to Appendix A, Figure \ref{fig:ViT}.}

    The Fourier neural operator (FNO) \cite{li2020fourier} model leverages the fast Fourier Transform to parameterize the integral kernel directly in the Fourier space. The implementation of FNO for the $2D$ Darcy Flow problem as presented in \cite{li2020fourier} is followed in this study. The total amount of parameters of FNO corresponds to 2.38 million, that is 15 times more than UNet+ViT (151k) and 17 times more than UNet (137k). 

\section*{Results}
    \subsection*{Forward problem with sparse observations}
    This section presents the prediction of the hydraulic head at sparse monitoring wells after a constant 10-day pumping period under two different training conditions. \textcolor{black}{We employ distinct sampling strategies for both input and output data in our methodology. Our training data is sampled from a regular quadratic grid, while for testing we have explored other arrangements, such as radial and helical, to understand their potential impact on the prediction performance.}
    In the first scenario, training is conducted using sparse data, with a spacing of 20 grid points for the input hydraulic conductivity field and a spacing of 8 for the output hydraulic head. Testing is then carried out on sparse data points, following the radial and helical patterns delineated in subsection~\ref{2.1}. The resulting root mean square error (RMSE) is found to be $ 5.2 \times 10^{-2}$,  $3.5 \times 10^{-2}$ and $8.1 \times 10^{-2}$ for the vanilla U-Net, the UNet+ViT models and FNO respectively. These results underline the superior performance of the UNet+ViT model in handling sparse data, exhibiting a lower RMSE compared to both the vanilla U-Net and the FNO models.
    
\begin{figure}[ht]
    \vskip 0.2in
    \begin{center}
    \centerline{\includegraphics[width=\columnwidth]{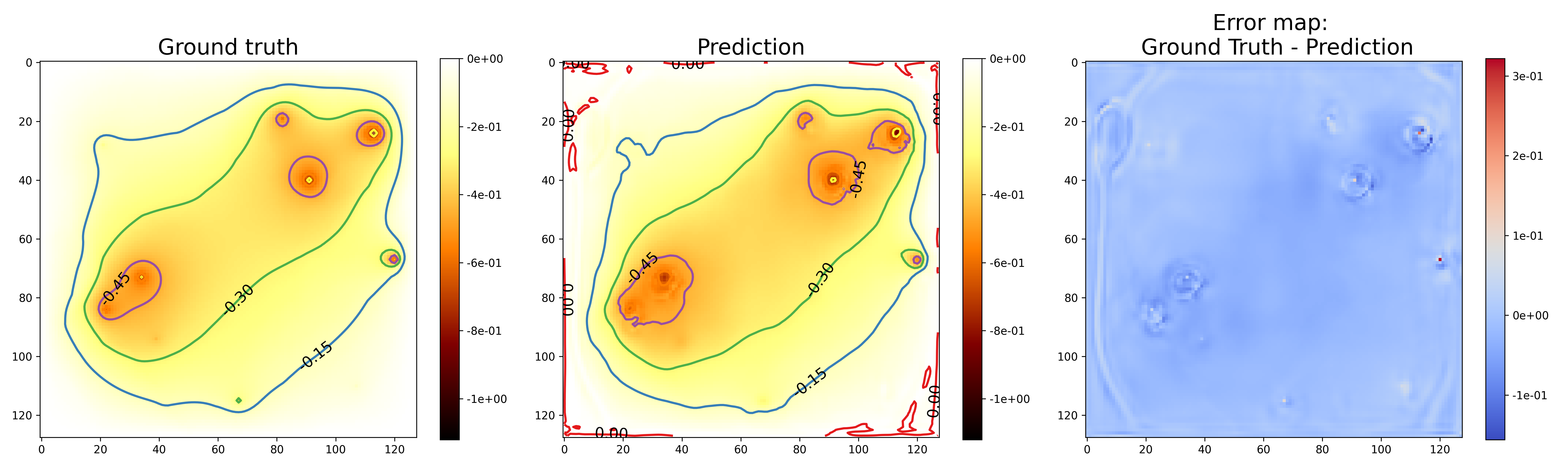}}
    \caption{Predictions over the entire domain are shown, with the UNet+ViT model exhibiting a RMSE of $1.9 \times 10^{-2}$. The figure demonstrates the model's ability to accurately capture the spatial distribution of the hydraulic head across the entire domain.}
    \label{forward-result}
    \end{center}
    \vskip -0.2in
    \end{figure}

    In contrast, when training is performed using the entire field and testing on the same sparse dataset, the error marginally escalates to $3.9 \times 10^{-1}$ for FNO, $3.8 \times 10^{-1}$ for UNet and $3.6 \times 10^{-1}$  for UNet+ViT model. This outcome is anticipated considering the training set exhibits sparsity in the first scenario, but not in the latter. Additionally, Figure~\ref{forward-result} displays the prediction over the entire domain, resulting in a lower RMSE of $1.0 \times 10^{-2}$ for FNO, $1.7 \times 10^{-2}$ and $1.9 \times 10^{-2}$ for the vanilla U-Net and UNet+ViT models, respectively. The FNO model, while superior when dealing with full data, exhibits the highest predictive error under sparse data observations. These results highlight the practical advantages of the U-Net and especially UNet+ViT model in real-world scenarios for which data sparsity is common.

     \textcolor{black}{It should be noted that traditional simpler neural networks and other machine learning techniques may not provide adequate solutions for this specific problem. This assertion is backed by a comparison of the results from a fully connected neural network, a linear regression model and a random forest, detailed in Appendix \ref{appendixB}. Despite the substantial number of trainable parameters, reaching 51.17 million, inherent to the fully connected neural network and the application of linear regression and random forest, these methods significantly underperform compared to the U-Net, the UNet+ViT models, and FNO.}

    \subsection*{Identification of pumping wells}
    In this section, we focus on an inverse problem: specifically the identification of pumping wells. This task requires determining the locations and rates of pumping wells based on the observed hydraulic heads. \textcolor{black}{Throughout these experiments, we employ a single hydraulic conductivity field, which, while spatially varying, remains identical across all samples within the dataset.}

    \textcolor{black}{In evaluating the performance of our models, we use both RMSE and accuracy. The RMSE calculates the average difference between the true and the predicted value for each pump location in the test dataset, giving a quantitative measure of the prediction error. Complementing this, the accuracy was determined by counting the proportion of correct pump predictions, where a prediction is considered correct if the predicted and actual pump locations align. This gives a sense of how often the model correctly identifies the location of pumps.}
    
    The U-Net model performs optimally, achieving an RMSE of $5.6 \times 10^{-2}$. Interestingly, the integration of the Vision Transformer with the U-Net model does not confer any additional precision in this scenario, yielding a near RMSE of $6.1 \times 10^{-2}$. The FNO model exhibits a higher RMSE of $1.1 \times 10^{-1}$, indicating a somewhat lower accuracy in identifying the pumping well locations.
    
\begin{figure}[ht]
    \vskip 0.2in
    \begin{center}
    \centerline{\includegraphics[width=0.8\columnwidth]{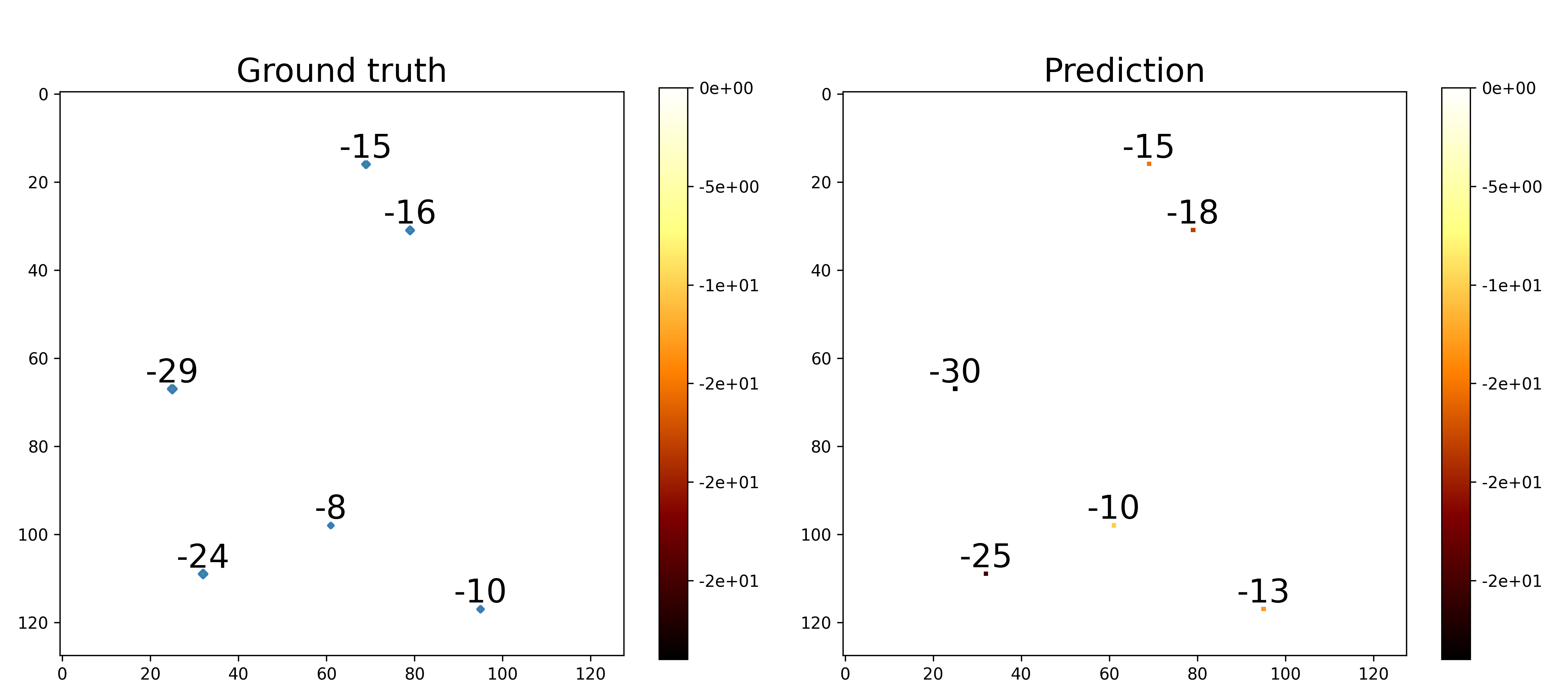}}
    \caption{Test sample results for the identification of pumping wells using the U-Net + ViT model. The model accurately identifies both the positions of the pumping wells with an accuracy of 93 \% across the entire test dataset.}
    \label{pumps-inverse-results}
    \end{center}
    \vskip -0.2in
    \end{figure}

    To visually illustrate these results, Figure~\ref{pumps-inverse-results} presents a test sample using the U-Net + ViT model. It demonstrates an accuracy of 93\% in locating the pumps, calculated across the entire test dataset. The figure visualizes the model's ability to accurately identify the positions and the pumping rate of the wells.  In comparison, the FNO model achieved a notably lower detection accuracy of 79\% in the same task.
    
    \subsection*{Example results for time series data}
\begin{figure}[ht]
    \vskip 0.2in
    \begin{center}
    \centerline{\includegraphics[width=0.8\columnwidth]{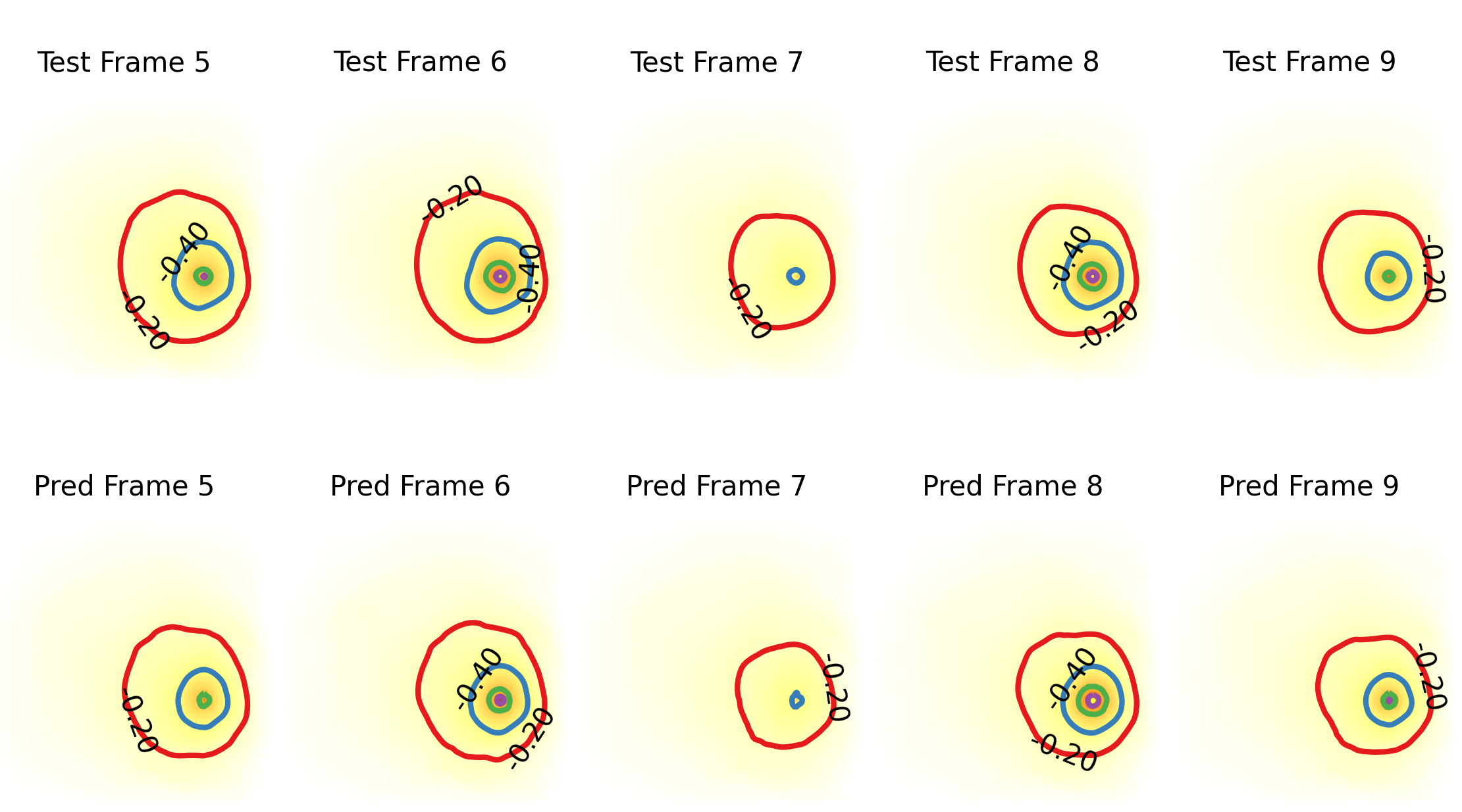}}
    \caption{\textcolor{black}{Depiction of autoregressive testing results from the UNet+ViT model over the final five frames of a ten-frame test sequence. This sequence simulates a single variable-rate pump operating over a ten-day period within a region characterized by a diverse hydraulic conductivity field. The top row presents the ground truth, while the lower row displays the predicted outcomes. Contour lines in the images represent groundwater levels, which are color-coded for enhanced visual clarity. Each successive frame is employed as input to the U-Net-Vision Transformer system, aiding in the prediction of the subsequent frame.}}
    \label{time-results}
    \end{center}
    \vskip -0.2in
    \end{figure}

    This section unveils the results achieved from the analysis of time series data, starting with a simplified scenario, for which the inputs are the varying hydraulic conductivity field and the pumping rate of a single pump which varies over a 10-day simulation period. Results are evaluated in terms of root mean square error (RMSE) with a focus on the comparison of different configurations of the U-Net architecture with transformers. 
    Figure~\ref{time-results} presents a comparison of results over 5 time frames for the U-Net with the Vision Transformer under autoregressive testing conditions.

    The RMSE for each method was calculated to quantify the models' performance. The U-Net architecture alone yielded an RMSE of $1.79 \times 10^{-2}$. When supplemented with a Vision Transformer, consisting of 2 attention blocks and 2 heads, the performance improves, registering an RMSE of $1.67 \times 10^{-2}$. However, increasing the complexity of the Vision Transformer to 8 blocks and 8 heads did not further improve the performance, instead, it led to a slight degradation in the RMSE ($1.77 \times 10^{-2}$). Adding an Axial Transformer \cite{ho2019axial} to the U-Net architecture also did not enhance the performance, yielding an RMSE of $1.83 \times 10^{-2}$.
    These results suggest that while adding a Vision Transformer to the U-Net architecture leads to performance improvement, increasing the complexity of the latent space does not necessarily do so.

\section*{Conclusion}

This paper explores and evaluates the capabilities of different machine learning models, with a particular focus on U-Net, U-Net integrated with Vision Transformers (ViT), and Fourier Neural Operator (FNO), in the context of predicting hydraulic head in groundwater studies.

Our analysis and testing, conducted on synthetic datasets designed to simulate the conditions from the Overbetuwe region in the Netherlands and including scenarios with sparse data, firmly establish that both U-Net and U-Net + ViT models are particularly adept at dealing with such tasks. Importantly, these models are also preferred due to their fewer requisite parameters.  
    
Specifically, in the case of sparse observation scenarios, the vanilla U-Net and the U-Net + ViT models outperformed the FNO model. In particular the performance of the UNet+ViT model was superior when handling sparse data, highlighting the potential of the model in real-world applications, where data scarcity is a common issue. The U-Net model demonstrated optimal performance in identifying pumping wells. Interestingly, the integration of the Vision Transformer with the U-Net model did not confer any additional accuracy in this scenario. As for the analysis of time series data, supplementing the U-Net architecture with a Vision Transformer improved the model performance, recording an RMSE of $1.67 \times 10^{-2}$ compare to $1.79 \times 10^{-2}$ of the vanilla U-Net. However, increasing the complexity of the Vision Transformer did not further enhance the model performance, indicating that a more complex architecture does not necessarily yield better results.

Future research will involve applying this validated methodology to real-world data, beginning with the Overbetuwe region in the Netherlands. This will offer an opportunity to further validate and refine the model, accounting for the sparsity and uncertainties inherent in real-world data.

\section*{Broader impact}
    
    The implications of this research span a wide range of potential societal impacts, with a primary focus on improving the efficiency and reliability of groundwater level forecasting. Given that groundwater is a crucial resource for approximately 2.5 billion people worldwide, fulfilling their daily water needs, and a significant source of global irrigation water, the importance of reliable forecasts cannot be overstated. Our work, through enhancing the performance of groundwater numerical models, offers an opportunity to revolutionize the management and distribution of this vital resource. By providing more accurate and data-efficient predictions, we can aid in the formulation of informed and sustainable water management strategies. This is particularly crucial considering the pressing challenges of population growth and climate change.

    \section*{Acknowledgements}
    \textcolor{black}{This work was carried out with support of the Leeds-York-Hull Natural Environment Research Council (NERC) Doctoral Training Partnership (DTP) Panorama under grant NE/S007458/1. 
    Our sincere appreciation is extended to Professor Karniadakis of Brown University. The  financial assistance provided by the Leeds Institute of Fluid Dynamics and Deltares, which made possible the research visit to Brown University, is also gratefully acknowledged. Lastly, we would like to express our gratitude to the reviewers. Their critiques and suggestions have greatly enhanced the overall clarity of our work.}


\newpage
\appendix
\onecolumn
\appendix

\section*{Appendix A} \label{appendixA}
    This appendix provides detailed diagrams of the model structures.

    \begin{figure}[ht]
    \centering
    \includegraphics[width=\columnwidth]{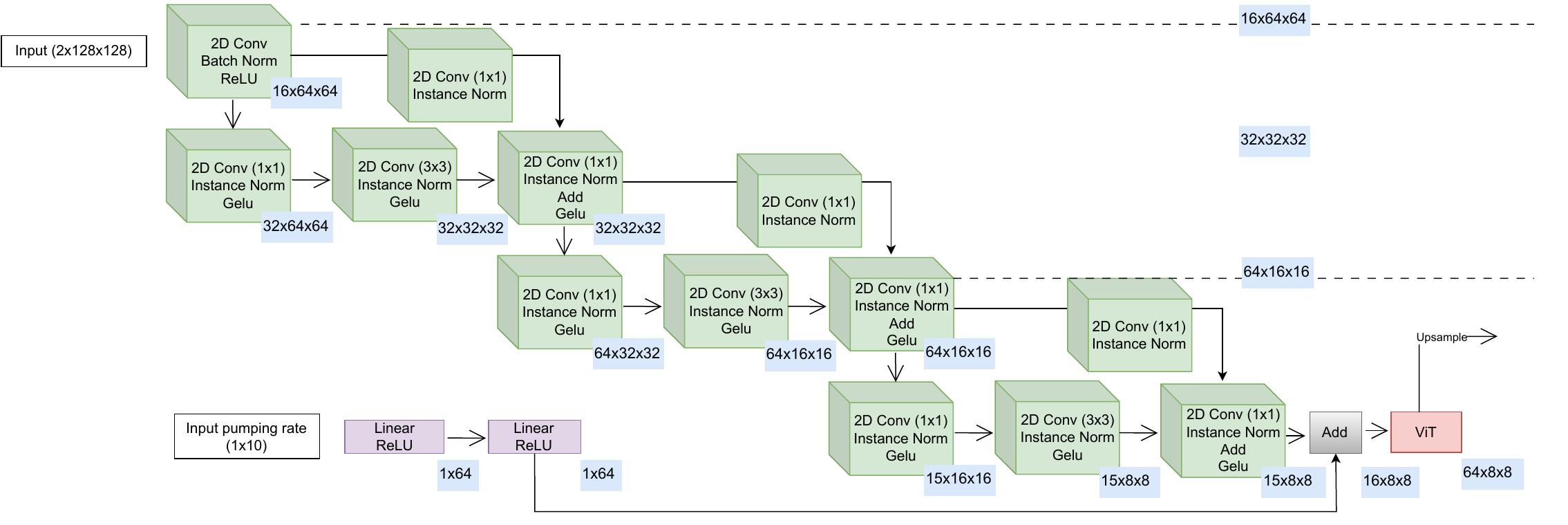}
    \vspace{-10pt}
    \caption{U-Net Encoder: The encoder consists of 3 bottleneck blocks, each comprising 3 convolutional blocks (Conv2d, Instance Normalization, and GELU). These blocks increase the channel dimension by a factor of 2 and perform downsampling with a stride of 2, extracting spatial features. In time-dependent scenarios, the pumping rate time series undergoes processing through a two-layer feed-forward neural network. The resulting processed data is then concatenated with the input, creating the latent space representation.}
    \label{fig:encoder}

    \includegraphics[width=\columnwidth]{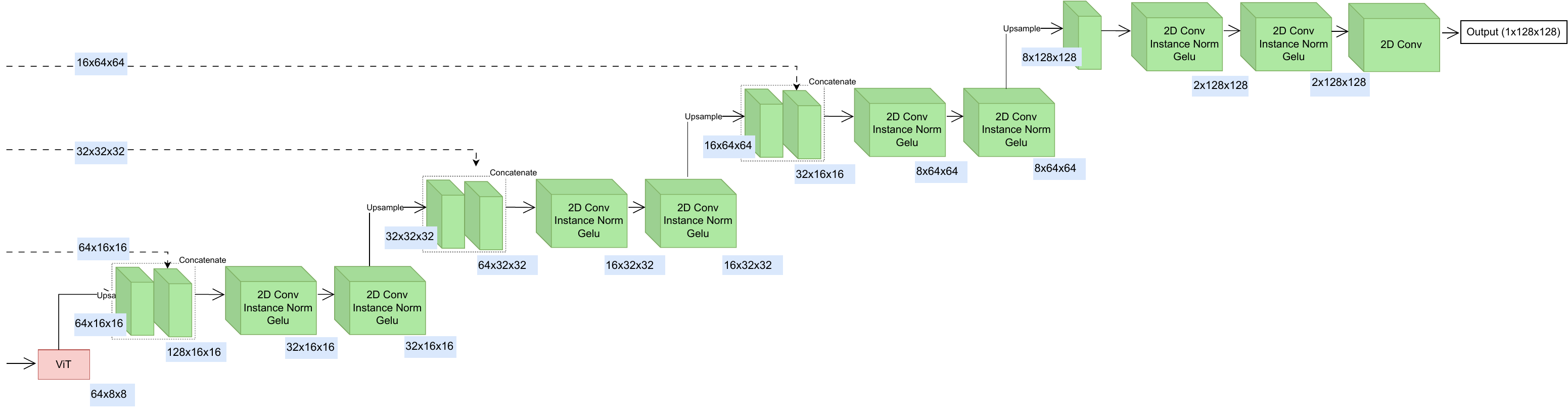}
    \caption{U-Net Decoder: The decoder is composed of a series of upsampling blocks, where each block consists of a bilinear upsampling operation followed by a double convolution operation.}
    \label{fig:decoder}

    \includegraphics[width=0.7\columnwidth]{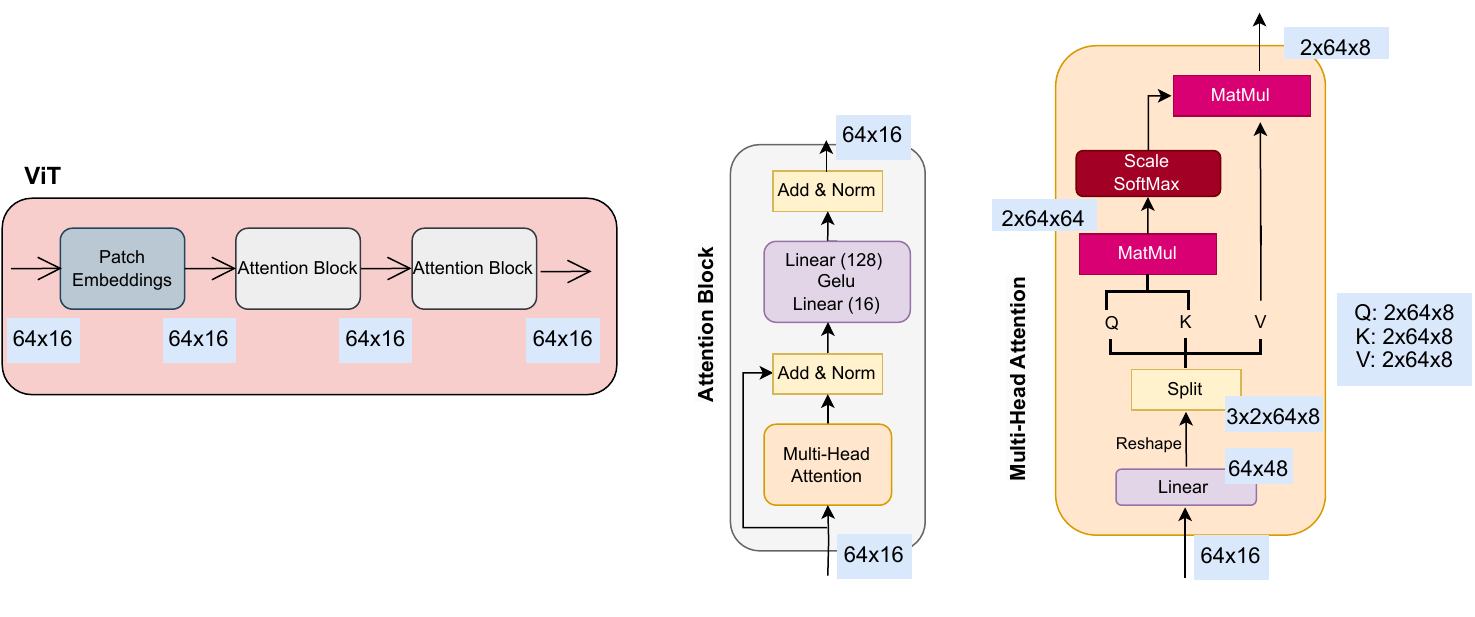}
    \caption{The Vision Transformer (ViT) and its components: The input is tokenized into a sequence of flattened 2D patches, each of size 1×1. The structure of the Transformer encoder includes 2 attention blocks, each containing Multi-Head Attention block and fully connected layers.}    \label{fig:ViT}
    \end{figure}

\section*{Appendix B} \label{appendixB}
    This appendix sets out to examine whether simpler machine learning models, specifically a fully connected neural network, a linear regression model, and a Random Forest model, can achieve the same level of accuracy as more advanced models like the U-Net, the UNet+ViT models, and FNO in predicting groundwater levels.

    The particular Random Forest model tested here used 30 estimators. The fully connected neural network, employed for this comparison, comprises three hidden layers, each containing 1000 nodes and using ReLU activation functions. The model holds an impressive count of 51.17 million trainable parameters. 
    
    Unfortunately, none of the models was able to predict accurately the groundwater levels neither capturing the location of the wells. Specifically, the fully connected neural network and the linear regression model yielded high RMSEs of $1.17 \times 10^{-1}$ and $1.24 \times 10^{-1}$, respectively. The Random Forest model fared slightly better, achieving a lower RMSE of $1.02 \times 10^{-1}$, but it still fell short of the U-Net, the UNet+ViT models, and FNO.
    
\begin{figure}[ht]
    \vskip 0.2in
    \begin{center}
    \centerline{\includegraphics[width=\columnwidth]{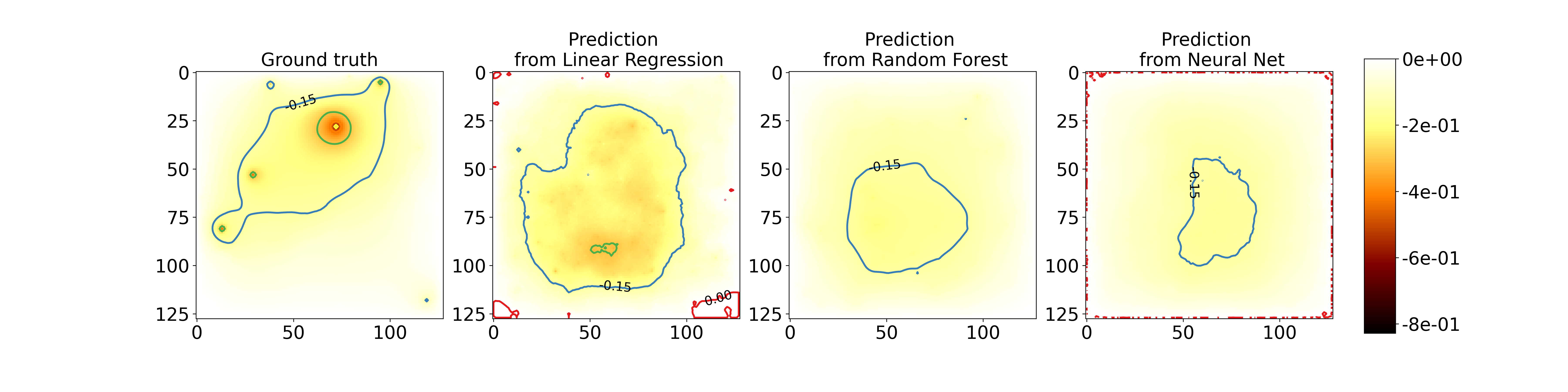}}
    \centerline{\includegraphics[width=\columnwidth]{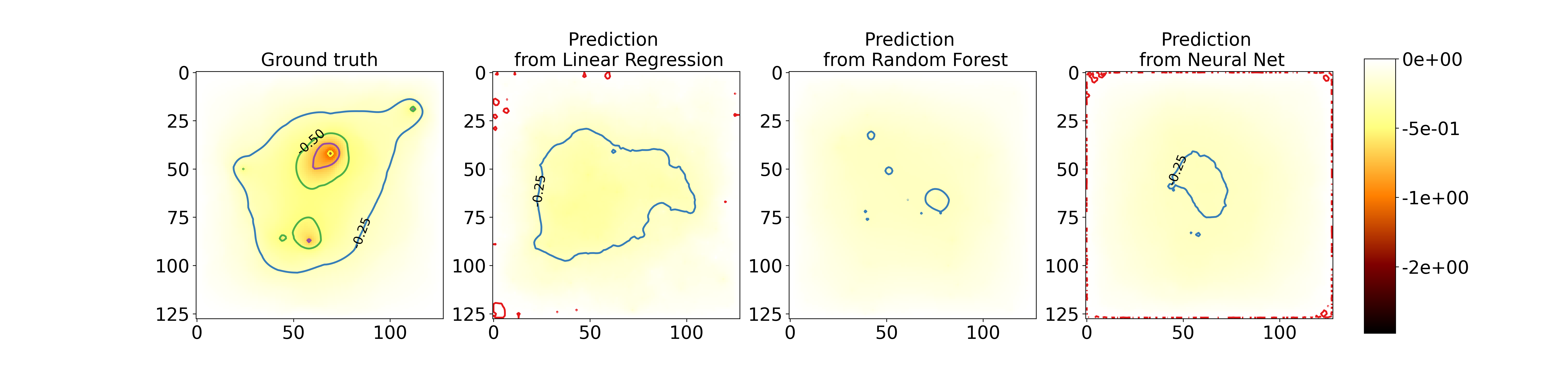}}
    \centerline{\includegraphics[width=\columnwidth]{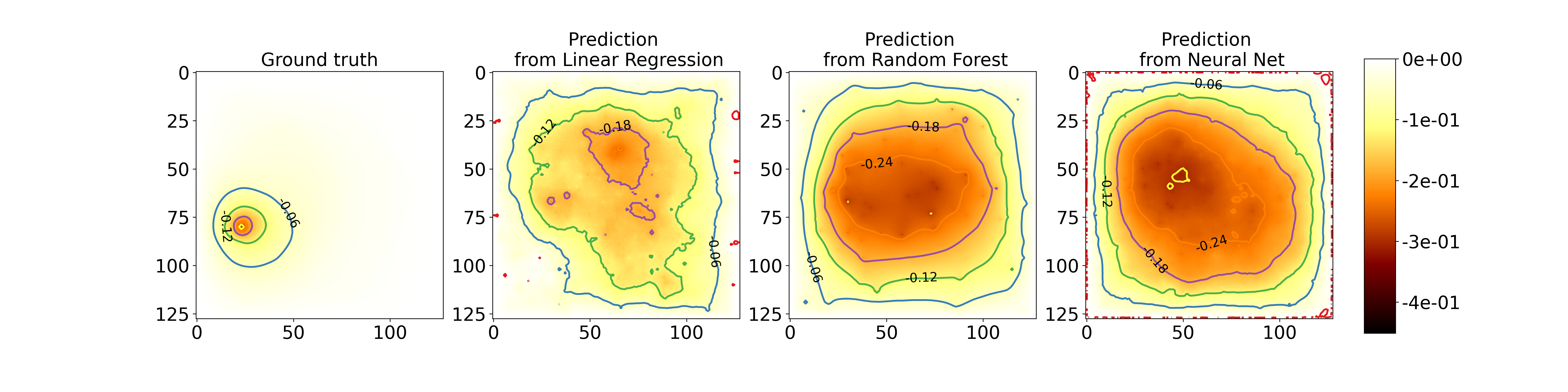}}
    \caption{These figures illustrate the testing results using a fully connected neural network, a linear regression model, and a Random Forest model. The top row of each figure presents the ground truth, while the remaining rows display the predicted outcomes generated by each respective model. Contour lines represent groundwater levels, color-coded for clarity.}
    \label{fig:comparison}
    \end{center}
    \vskip -0.2in
    \end{figure}

    Figure~\ref{fig:comparison} visually contrasts the predictions of these simpler models gainst the ground truth. Their significant underperformance becomes evident when compared to more sophisticated models. For a comparison of these results with accurate outcomes produced by the UNet+ViT model, the reader is directed to Figure~\ref{forward-result}.

\end{document}